\documentclass[twocolumn]{aastex62}

\usepackage[caption=false]{subfig}
\usepackage{float}
\usepackage{graphicx}	
\usepackage{amsmath}	
\usepackage{amssymb}
\usepackage{xcolor}

\graphicspath{{./}}

\received{24 April 2020}
\accepted{}

\shorttitle{Cr and V in ultra hot jupiter WASP-121~b}
\shortauthors{Ben-Yami et al.}

\begin{document}

\title{Neutral Cr and V in the atmosphere of ultra hot jupiter WASP-121~b}

\correspondingauthor{Nikku Madhusudhan}
\email{nmadhu@ast.cam.ac.uk}

\author{Maya Ben-Yami}
\affil{Institute of Astronomy, University of Cambridge, Madingley Road, Cambridge CB3 0HA, UK}

\author{Nikku Madhusudhan}
\affil{Institute of Astronomy, University of Cambridge, Madingley Road, Cambridge CB3 0HA, UK}

\author{Samuel H. C. Cabot}
\affil{Yale University, 52 Hillhouse, New Haven, CT 06511, USA}

\author{Savvas Constantinou}
\affil{Institute of Astronomy, University of Cambridge, Madingley Road, Cambridge CB3 0HA, UK}

\author{Anjali Piette}
\affil{Institute of Astronomy, University of Cambridge, Madingley Road, Cambridge CB3 0HA, UK}

\author{Siddharth Gandhi}
\affil{Department of Physics, University of Warwick, Gibbet Hill Road, Coventry CV4 7AL, UK}
\affil{Centre for Exoplanets and Habitability, University of Warwick, Gibbet Hill Road, Coventry CV4 7AL, UK}

\author{Luis Welbanks}
\affil{Institute of Astronomy, University of Cambridge, Madingley Road, Cambridge CB3 0HA, UK} 



\begin{abstract}
Ultra hot jupiters (UHJs), giant exoplanets with equilibrium temperatures above 2000 K, are ideal laboratories for studying metal compositions of planetary atmospheres. At these temperatures the thermal dissociation of metal-rich molecules into their constituent elements makes these atmospheres conducive for elemental characterisation. Several elements, mostly ionized metals, have been detected in UHJs recently using high resolution transit spectroscopy. Even though a number of neutral transition metals (e.g., Fe, Ti, V, Cr) are expected to be strong sources of optical/NUV opacity and, hence, influence radiative processes in the lower atmospheres of UHJs, only Fe~I has been detected to date. We conduct a systematic search for atomic species in the UHJ WASP-121~b. Using theoretical models we present a metric to predict the atomic species likely to be detectable in such planets with high resolution transmission spectroscopy. We search for the predicted species in observations of WASP-121~b and report the first detections of neutral transition metals Cr~I and V~I in an exoplanet at 3.6 $\sigma$ and 4.5 $\sigma$, respectively. We confirm previous detections of Fe~I and Fe~II. Whereas Fe~II was detected previously in the NUV, we detect it in the optical. We infer that the neutral elements Fe~I, V~I, and Cr~I are present in the lower atmosphere, as predicted by thermochemical equilibrium, while  Fe~II is a result of photoionisation in the upper atmosphere. Our study highlights the rich chemical diversity of UHJs. 
\end{abstract}



\section{Introduction} \label{sec:intro}
Ultra Hot Jupiters (UHJs) are emerging as promising laboratories for exploring the metal compositions of giant exoplanets. Due to their close proximity to their host stars these extremely irradiated gas giants have atmospheric temperatures of $\sim$2000-4000 K, providing a unique opportunity to study planetary processes under extreme conditions, e.g., atmospheric escape \citep{Ehrenreich2015,Yan2018}, day-night chemical variations  \citep{Wong2019, Ehrenreich2020} and thermal inversions \citep{Haynes2015,Evans2017}. Additionally, the molecules in such atmospheres can be thermally dissociated into atoms and ions that can be observed using high resolution spectra \citep{Kitzmann2018, Lothringer2018, Hoeijmakers2019}.

In recent years, several metallic species have been detected in UHJs, predominantly through high resolution Doppler spectroscopy \citep[e.g.,][]{Hoeijmakers2019,CasasayasBarris2019}. The detections are typically made using the cross-correlation method \citep{Snellen2010}, which has proven to be a robust way to detect molecular, atomic and ionic species in hot Jupiters in general \citep{Snellen2010, Brogi2012, Birkby2018, Hoeijmakers2019}. The elemental detections include Na~I, Mg~I, Sc~II, Ti~II, Ca~II, Cr~II, Fe~I, Fe~II and Y~II in KELT-9 b \citep{Hoeijmakers2018, Hoeijmakers2019, Cauley2019, Yan2019, Turner2020}; Na~I, Ca~II, Mg~I, Cr~II, Fe~I and Fe~II in MASCARA-2 b \citep{CasasayasBarris2019, Strangret2020, Nugroho2020, Hoeijmakers2020}; Ca~II in WASP-33b \citep{Yan2019}; and Na~I, Fe~I, Fe~II, Mg~II in WASP-121~b \citep{Sing2019,Bourrier2020, Cabot2020, Gibson2020}. 

Most of the metal detections in UHJs to date are ions, with the exception of prominent species Na~I, Mg~I and Fe~I, all of which have solar abundances above 1 ppm. On the other hand, the lower atmospheres of UHJs are expected to host a number of trace metals in neutral form \citep{Kitzmann2018}. In particular, trace transition metals (e.g., Ti~I, V~I, Cr~I, Mn~I) can be strong sources of optical/NUV opacity \citep{Lothringer2020} and can provide important probes of physicochemical processes in the lower atmosphere, as discussed in Section~\ref{sec:discussion}.

Therefore, a consistent framework to determine a priori the detectability of atomic species would be valuable to inform observations and allow for comparative chemical characterisation of UHJs. We explore such an approach in the present work using the case study of WASP-121~b, an UHJ of considerable interest from recent studies \citep[e.g.][]{Delrez2016, Evans2017, Evans2018, Bourrier2020, Parmentier2018}. Notably, it has been suggested to contain a thermal inversion \citep{Evans2017}. While TiO/VO are traditionally considered strong candidates for causing thermal inversions \citep{Fortney2008} they have proved elusive in this planet \citep{Evans2018, Mikal-Evans2019, Merritt2020}, prompting considerations of other species \citep{Lothringer2018,Gandhi2019}. Furthermore, its bright host star ($V=10.44$) and inflated radius make WASP-121~b particularly amenable to atmospheric characterization. We consider a metric to predict the detectable atomic species in WASP-121~b based on their abundances in thermochemical equilibrium and the number and strength of their spectral lines. We then search for the predicted species in high resolution optical spectra of WASP-121~b, and confirm our predictions. 

\section{Modelling and Theory} \label{sec:model}
\begin{figure*}[htb!]
    \centering
        \subfloat{\includegraphics[width=0.48\textwidth]{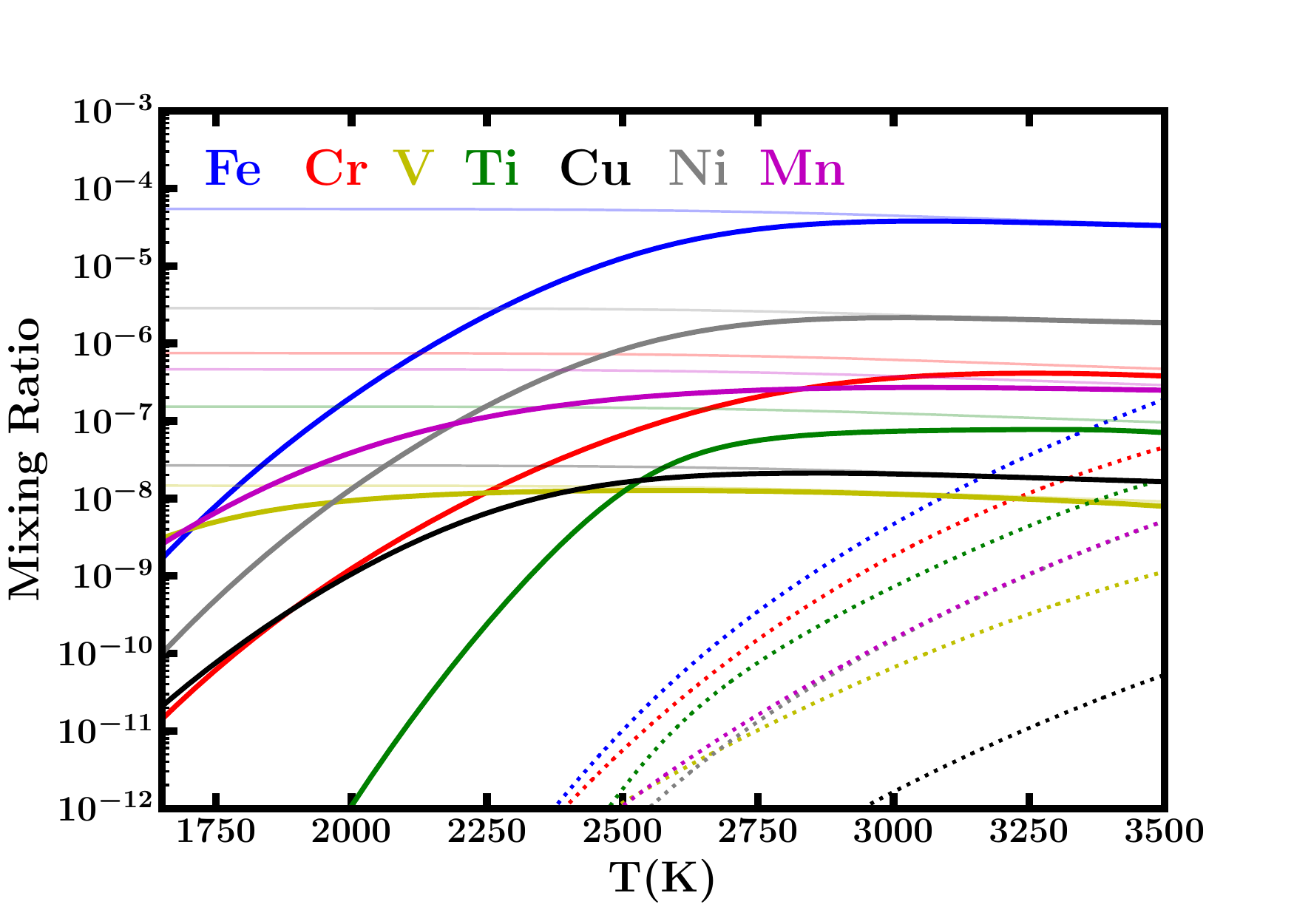}}
        \subfloat{\includegraphics[width=0.48\textwidth]{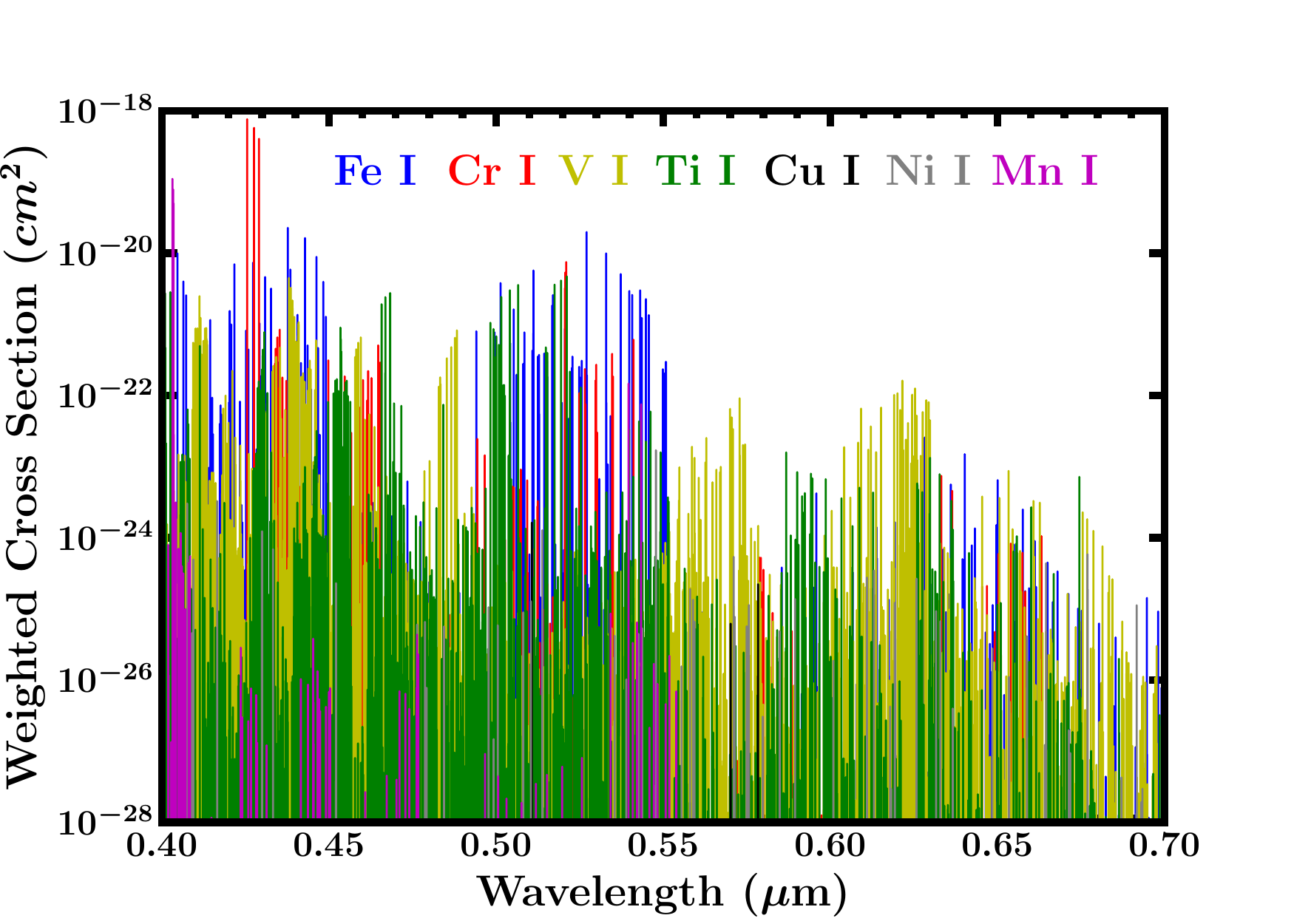}}
    \caption{Equilibrium volume mixing ratio and absorption cross sections for several transition metals. \textit{Left}: The solid (dotted) lines show the neutral (singly ionised) abundances as a function of temperature at 0.1 bar pressure assuming solar elemental abundances \citep{Asplund2009}; see section~\ref{subsec:op}. The thin solid line for each species shows the mixing ratio it would have if all its elemental abundance was in neutral atomic form. \textit{Right} : Absorption cross sections at 2000 K weighted by their solar elemental abundances.}
    \label{fig:chemeq}
\end{figure*}

Here we explore the expected atomic composition and spectral signatures of UHJ atmospheres with WASP-121~b as our case study. We asses the prominent species with thermochemical equilibrium calculations, compute absorption cross-sections for these species and use these cross-sections to investigate optical transmission spectra of WASP-121~b.
\newpage
\subsection{Atomic Abundances} \label{subsec:op}
We examine the atomic abundances in the photospheres of UHJs under the assumption of thermochemical equilibrium. The chemical  calculations were conducted using the HSC \textsc{Chemistry} software (version 8)\footnote{\url{www.hsc-chemistry.com}}. We assumed solar elemental abundances \citep{Asplund2009}. The input species were expanded upon those used by \citet{Harrison2018} to include important molecular, atomic and ionic forms of the top 36 elements in solar abundance. Other similar calculations show the prominence of atomic species in UHJs \citep[e.g.,][]{Kitzmann2018,Lothringer2018,Nugroho2020}.

Although the equilibrium temperature of WASP-121b is $\sim$2400 K, we expect its observed composition to be characteristic of the ~2800 K temperatures at $\sim$0.1-1 bar \citep{Gandhi2019} due to strong vertical mixing \citep{Parmentier2013}. Figure \ref{fig:chemeq} shows abundances of some prominent transition metals as a function of temperature at 0.1 bar. It can be seen that above $\sim$2500~K these elements are primarily found in their neutral atomic form rather than in molecules. We thus take the total elemental abundance, which we have assumed to be solar, as a good approximation for the abundance of neutral atomic species above 2500~K.

\subsection{Atomic Cross sections} \label{subsec:xsec}
We calculate atomic absorption cross sections for a large number of elements in order to investigate their spectral features. We rank the elements by their solar abundances \citep{Asplund2009}, excluding H, noble gases and halogens. We calculate the cross sections for all the neutral and singly ionised forms of the top 30 elements, following the methods of \citet{Gandhi2017}. We use atomic line lists provided by \citet{Kurucz2018} similar to the approach in other recent studies \citep[e.g.,][]{Hoeijmakers2018,CasasayasBarris2019, Nugroho2020}. We determine the partition functions from the energy levels and statistical weights from the NIST database \citep{Kramida2018}. We include the effect of thermal broadening of the atomic lines, but do not consider pressure broadening as high resolution spectra probe lower pressure regions in the upper atmosphere \citep{Wyttenbach2015, Snellen2010}. Our calculations do not include the effect of rotation and atmospheric expansion on the shape of the lines.
All our models also include Rayleigh scattering from H$_2$ and collision-induces-absorption (CIA) from H$_2$-H$_2$ and H$_2$-He \citep{Richard2012}. An example of these cross sections scaled by solar abundance for some prominent transition metals can be found in Figure \ref{fig:chemeq}.

\subsection{Theoretical Spectra} \label{subsec:theo}

\begin{figure*}[htb]
    \centering
        \subfloat{\includegraphics[width=0.98\textwidth]{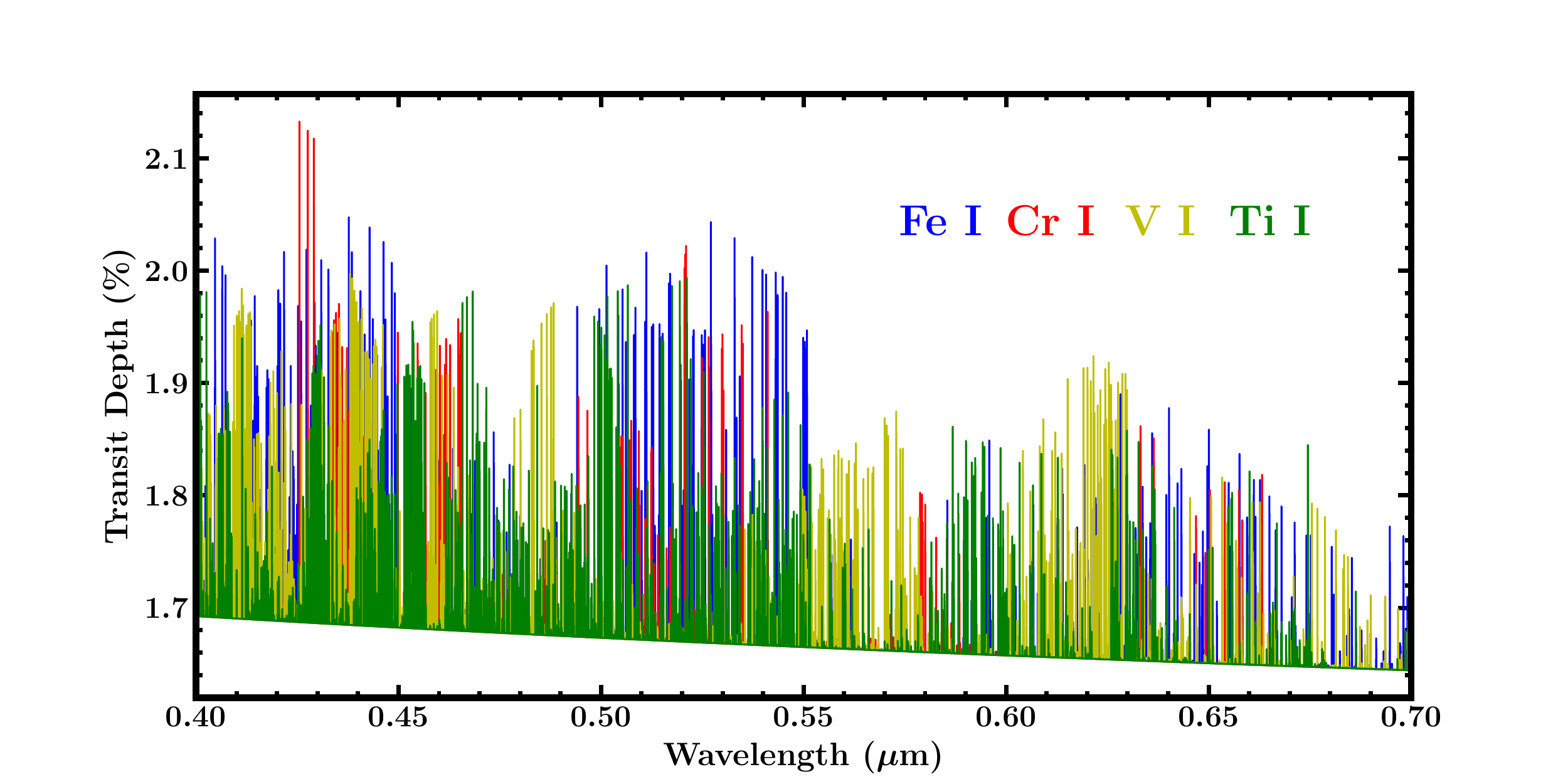}}
    \caption{Model transmission spectra of WASP-121~b with contributions from Fe~I, Cr~I, V~I and Ti~I. See section~\ref{subsec:theo}.}
    \label{fig:analytic}
\end{figure*}

We investigate the spectral signatures of the atomic species discussed above using both simple semi-analytic models as well as fully numerical models. First, we compute transmission spectra of WASP-121~b using the analytic formulation originally derived by \citet{Lecavelier2008} \citep[also see e.g.,][]{DeWit2013,Betremieux2017}. This prescription assumes an isothermal atmosphere with a uniform composition and a constant scale height. The wavelength dependent effective height of the atmosphere is given by:
\begin{equation}
    h(\lambda) =H\ln{\left(\frac{\sqrt{2\pi R_{pl}H}n_0}{\tau_{eq}}\sigma_\text{total}\right)},
    \label{eq:h_final}
\end{equation}
with total opacity
\begin{equation}
    \sigma_\text{total} = \sum_i \chi_i\sigma_i(\lambda)  + \sigma_R(\lambda) + n_0\sigma_\text{CIA}(\lambda),
    \label{eq:total_op}
\end{equation}
where $\sigma_R$ is the opacity due to Rayleigh scattering, $\sigma_i$ and $\chi_i$ are the absorption cross section and mixing ratio of species $i$, and $n_0\sigma_\text{CIA}$ is the appropriately scaled CIA cross section. The CIA approximation in Equation \ref{eq:total_op} is adequate here considering that the CIA contribution in the optical is negligible. The scale height is given by $H=(k_BT)/(\mu g)$ where $k_B$ is the Boltzmann constant, $T$ is the temperature, $\mu$ is the mean molecular mass and $g$ is the gravitational acceleration. $R_{pl}$ is the planet continuum radius, $n_0$ is the number density at a given reference pressure $P_0$, and $\tau_{eq} = 0.56$ \citep{Lecavelier2008}. For a star of radius $R_\text{star}$ the absorption spectrum is then:
\begin{equation}
    \delta(\lambda) =\frac{\left(R_\text{pl}+h(\lambda)\right)^2}{R_\text{star}^2}
    \label{eq:td}
\end{equation}
We generate such analytic spectra for atomic species in the atmosphere of the ultra hot Jupiter WASP-121~b. We consider the contribution to the opacity of only one species $j$ at a time, in addition to continuum opacity due to Rayleigh Scattering and CIA, to calculate the spectrum $\delta_\text{$c$+j}$. Here $\chi_j$ is taken to be the solar elemental mixing ratio of the species $j$. The corresponding atomic cross sections are calculated at 2000~K as described in Section \ref{subsec:xsec}. Examples for Fe~I, Cr~I, V~I and Ti~I can be seen in Figure \ref{fig:analytic}. We adopt the planet parameters from \cite{Delrez2016} as $R_p = 1.828R_J, M_p=1.183M_J, R_{star}=1.458R_{sun}$. We assume $\mu=2.22$ (similar to Jupiter), $P_0=0.1$ bar and $T=2000$~K, and calculate the spectra from 0.4 to 0.7$\mu$m with a resolution of R = $10^5$. We use these analytic model spectra to compute our detectability metric for the different species, as discussed in section~\ref{subsec:predictions}.

For the cross-correlation analyses we use fully numerical models to compute the transmission spectra for the species discussed above. We use an adapted version of the forward model in the AURA retrieval code \citep{Pinhas2018}. AURA computes line-by-line radiative transfer for a plane-parallel atmosphere in transmission geometry, assuming hydrostatic equilibrium. We assume uniform chemical abundances and an isothermal temperature profile at 2000 K. The sources of opacity and spectral range are the same as described above for the analytic models. We compare the numerical models with the analytic models discussed above and find them to be in good agreement.

\section{Observations and Analysis} \label{sec:obs}

We demonstrate our predictive methodology on archival transit observations of WASP-121~b. The observations were conducted using the High Accuracy Radial Velocity Planet Searcher (HARPS) spectrograph installed at the ESO La Silla 3.6m telescope, as part of observing programme 0100.C-0750(C) (Hot Exoplanet Atmospheres Resolved with Transit Spectroscopy, PI: D. Ehrenreich). Three transits were observed on nights of 31 December 2017, 09 January 2018, and 14 January 2018 (hereafter Nights 1, 2 and 3). Data were reduced with HARPS DRS v3.8. The extracted 68 orders span 380-690nm at $R\sim115,000$. Spectra acquired on Night 2 have low signal-to-noise ratio (likely from poor seeing conditions), and are excluded from our analysis. We treat data from Nights 1 and 3 separately in our analysis until computation of detection significances.

\subsection{Detrending and Cross-correlation}

\begin{table*}[t!]
\centering
\begin{tabular}{lllllllllllllll}
{}                 & {Fe I} & {Ti I} & {V I} & {Cr I} & {Sc I} & {Na I} & {Y I} & {Ca I} & {K I} & {Zr I} & {Mn I} & {Sr I} & {Mg I} & {Ni I} \\ \hline
{$N_\text{lines}$} & 185           & 156           & 149         & 71           & 29           & 8            & 30          & 22           & 9           & 28           & 12           & 4            & 2            & 1            \\
{S/N$_\text{av}$}  & 1.4           & 1.28          & 1.29        & 1.4          & 1.32         & 2.44         & 1.25        & 1.39         & 2.06        & 1.07         & 1.57         & 1.38         & 1.74         & 1.18          \\ \hline
{$\Psi$}           & 19.02         & 16.03         & 15.78       & 11.81        & 7.12         & 6.91         & 6.85        & 6.5          & 6.17        & 5.68         & 5.45         & 2.75         & 2.45         & 1.18       
\end{tabular}
\caption{Detectability metric ($\Psi$) with $\xi=2000$ ppm for various neutral atomic species with strong lines in the optical as predicted for the transmission spectrum of WASP-121~b (see section~\ref{sec:results}).}
\label{tb:lines}
\end{table*}

The majority of our analysis involves common steps in high resolution atmospheric spectroscopy \citep{Snellen2010, Brogi2012, Birkby2018, Hoeijmakers2019, CasasayasBarris2019}. We use the \texttt{X-COR} cross-correlation software suite, previously used for molecular detections \citep{Hawker2018, Cabot2019} in the near-infrared, and atomic detections on the present dataset \citep{Cabot2020}. We briefly describe our procedures here as follows; further details can be found in \citet{Cabot2020}. 

We divide the data by a telluric model, which is fitted and computed by \texttt{molecfit} v1.5.7 \citep{smette2015}. We mask the HARPS chip gap, plus $1\%$ of data from either end of the observed spectrum which are otherwise affected by low throughput or severe telluric contamination.
We shift all spectra into the rest-frame of the star by correcting for the planetary-induced radial velocity signal. Next, we construct a master stellar spectrum by co-adding all out-of-transit spectra. Each in-transit spectrum is then divided by the master stellar spectrum. We fit and divide each spectrum by a $5^{\rm th}$-degree polynomial, and remove remaining broadband variation with a 75-pixel high-pass filter. We are left with individual transmission spectra (Equation~\ref{eq:td}) which contain spectral features originating in the planet's atmosphere. Continuum information was removed in the above normalization.

We form cross-correlation templates by subtracting the continua from the numerical model transmission spectra computed in Section~\ref{subsec:theo}. We convolve the templates with a narrow Gaussian (FWHM$\sim$0.8 km s$^{-1}$) to approximate the HARPS spectral resolution. The Cross-Correlation Function (CCF) between the model template $m(\lambda)$ and observed transmission spectra $f(\lambda, t)$ is defined as:
\begin{equation}
    {\rm CCF}(v, t)=
    \frac{\sum_\lambda m(\lambda; v)f(\lambda, t)/\sigma^2(\lambda)}
    {\sum_\lambda m(\lambda; v)/\sigma^2(\lambda)}, 
    \label{eq:cc}
\end{equation}
where we weight summation over wavelength bins $\lambda$ by the inverse of their time-axis variance $\sigma^2(\lambda)$. This step prevents noisy data (e.g. in the former cores of telluric or stellar lines) from dominating the dot-product. The model is Doppler shifted by velocities $-600 \leq v \leq 600$ km s$^{-1}$ in steps of 2 km s$^{-1}$. The planetary atmospheric signal manifests as a light trail in the CCF, tracing velocities which correspond to the radial component of its orbital motion (which varies by $\sim100$ km s$^{-1}$ throughout the course of the transit). For species in common with the host star, we also observe a dark trail due to the Rossiter-McLaughlin effect \citep{Cegla2016}; a result of stellar rotation and the planet selectively occulting portions of the star. We correct for the Doppler Shadow as in \citet{Cabot2020}. Following \citet{Brogi2012}, we stack the CCFs in time for various values of semi-amplitude ${K_{p}}$. Excess signal from stacking under the true ${K_{p}}$, offset by the true systemic velocity $V_{\rm sys}$, constitutes an atmospheric detection. The final CCFs combine data from Nights 1 and 3.

\begin{figure*}
    \centering
    \includegraphics[width=\linewidth]{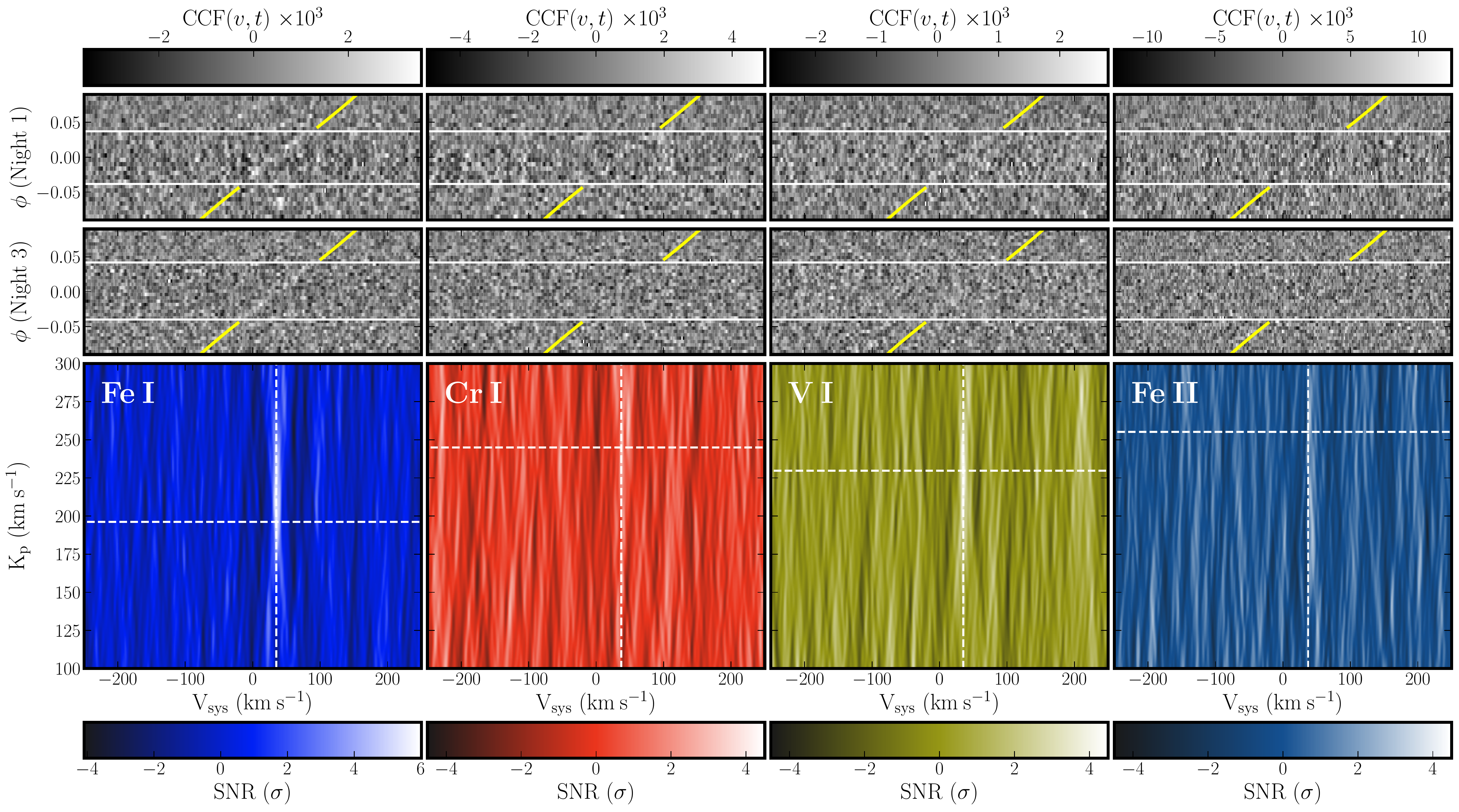}
    \caption{Detection of chemical species in WASP-121~b. {\it Top gray panels}: CCF in orbital phase and velocity. Horizontal white lines mark the transition between out-of-transit and in-transit. A yellow solid line shows the expected radial velocity of the planet in the out-of-transit frames. {\it Bottom color panels}: CCF in the $K_p$-$V_{\rm sys}$ plane for different species; see section~\ref{sec:results}. Dashed white lines indicate the peak CCF value. We report detections of Cr~I and V~I, and confirm previous detections of Fe~I and Fe~II.}
    \label{fig:results1}
\end{figure*}

\section{Results} \label{sec:results}
We now use our model spectra to predict the atomic species detectable in the atmosphere of WASP-121~b and then search for them in observed high resolution spectra. We first discuss our model predictions and then present new detections of Cr~I and V~I, besides confirming previous detections of Fe~I and Fe II. 
\subsection{Model Predictions} \label{subsec:predictions}
Here we use a simple metric to predict the detectability of a given atomic species using high resolution optical transmission spectra of a UHJ. Our metric is guided by the fact that detecting $N_\text{lines}$ lines using the cross-correlation method boosts the signal-to-noise ratio (S/N) by $\sqrt{N_\text{lines}}$ \citep{Birkby2018}. These lines are those that are strong enough to be detectable. The contribution of species $j$ to the spectrum is given by: 
\begin{equation*}
    \Delta \delta_\text{j} = \delta_\text{c+j} - \delta_\text{c},
\end{equation*}
where $\delta_\text{c}$ and $\delta_\text{c+j}$ are model transmission spectra with just the continuum opacity and that with contribution from the species $j$, respectively (see Section \ref{sec:model}).

We calculate the number of strong lines $N_\text{lines}$ in this signal by choosing a noise level $\xi$ and discarding all lines weaker than $\xi$. The S/N for a line is given by $\Delta \delta_j/\xi$ for that line. We take into account the strength of these $N_\text{lines}$ lines by calculating their average S/N, denoted as S/N$_\text{av}$. Finally, our metric for assessing the detectability is given by: $\Psi = \sqrt{N_\text{lines}}\cdot \text{S/N}_\text{av}$. 

For WASP-121~b, we use this method and the analytic model spectra of 30 neutral atomic species, as described in Section \ref{sec:model}, to evaluate their detectability. Here, we adopt a conservative noise-level of $\xi$=2000 ppm at R = 10$^5$. We present the metric for all species with strong lines above the continuum in Table \ref{tb:lines}. Following these results, in our analysis of observed spectra of WASP-121~b we search for all the species with $N_\text{lines}>25$: Fe~I, Ti~I, V~I, Cr~I, Sc~I, Y~I and Zr~I. These are the species we predict to be likely detectable in the transmission spectra of WASP-121~b, with Fe~I, Ti~I, V~I, and Cr~I being the four most likely. We note that this metric is only applicable to the species we expect to find under equilibrium conditions, typically seen in the lower atmosphere.

\begin{table}
  \centering
  \begin{tabular}{@{}l c c c r@{}}
   \hline
   \hline
   Species & S/N & Weighted Absorption & $K_{p}$ & $V_{\rm sys}$ \\
   \hline
    Fe~\textsc{I}   & 6.0 & 0.11\% & $196_{-11}^{+35}$ & $35_{-1}^{+2}$ \\
    Cr~\textsc{I}   & 3.6 & 0.10\% & $245_{-18}^{+15}$ & $37_{-2}^{+2}$ \\
    V~\textsc{I}    & 4.5 & 0.08\% & $229_{-23}^{+20}$ & $35_{-3}^{+2}$ \\
    Fe~\textsc{II}  & 3.6 & 0.23\% & $255_{-13}^{+6}$ & $37_{-1}^{+1}$ \\
   \hline
  \end{tabular}
  \caption{Summary of species detected in the atmosphere of WASP-121~b. The S/N corresponds to the peak of CCF in the $K_p$-$V_{\rm sys}$ plane. The weighted absorption provides a measure of the average line strength.}
  \label{tb:det}
\end{table}

\subsection{Chemical Detections}
Based on the predictions above we search for neutral and singly ionised Fe, V, Cr, Ti, Sc, Y and Zr in WASP-121~b using high resolution optical transmission spectra as discussed in section \ref{sec:obs}. The detections from our cross-correlation analyses are shown in Figure \ref{fig:results1}. We report new detections of V~I with 4.5$\sigma$ and Cr~I with 3.6$\sigma$ significance, and confirm the previous detection of neutral Fe at 6.0$\sigma$ significance \citep{Bourrier2020,Cabot2020, Gibson2020}. We find no significant excess signal in the cross-correlation functions of Ti~I, Sc~I, Y~I or Zr~I. We also report a detection of Fe~II in the optical, confirming its previous detection by \citet{Sing2019} in the UV. The weights in the CCF in Equation~\ref{eq:cc} are not optimized for model injection and recovery \citep{Hoeijmakers2019}. As such, our quoted detection significances represent conservative estimates.

The $K_p$ and $V_{\rm sys}$ values derived from the detected planetary absorption signals are consistent, within $1-2\sigma$, with the true values of ${K_{p}}=217\pm19$ km s$^{-1}$ and $V_{\rm sys}=38.350\pm0.021$ km s$^{-1}$ \citep{Delrez2016}. Our measurements of $K_p$ and $V_{\rm sys}$ are listed in Table~\ref{tb:det}. Our detections of neutral species exhibit blueshifts between $1-3$ km s$^{-1}$ at a $\lesssim 2\sigma$ level. This may be explained by wind speeds in the photosphere, as inferred in other hot Jupiter atmospheres \citep[e.g.,][]{Louden2015} and predicted by models of atmospheric dynamics \citep{Showman2009, Kempton2012}. The large spread in $K_p$ results from sampling a small portion of the planet's full orbit \citep{Brogi2018}.

We estimate the neutral Fe, Cr and V to pertain to lower regions in the atmosphere relative to Fe~II. The weighted absorption \citep{Hoeijmakers2019} by neutral species corresponds to $\sim 1.02$ to $1.04 R_p$, whereas Fe~II extends to $\sim 1.07 R_p$. For a scale height of $\sim$1000 km \citep{Evans2018}, and nominal pressure of 0.1 bar at the base of the atmosphere \citep{Welbanks2019}, the neutral species lie at altitudes corresponding to $3\times10^{-3}$ to $9\times10^{-4}$ bar. For comparison, the stratosphere \citep{Evans2017} corresponds to approximately $10^{-2}$ to $10^{-4}$ bar. On the other hand, we infer Fe~II to be present in the upper atmosphere ($\sim 6\times10^{-6}$ bar) suggesting it originates in photoionisation of Fe~I in agreement with \citet{Sing2019}.

The detections we report and the order of their significances are consistent with our predictions. However, we do not detect Ti~I even though its predicted detectability ($\Psi$) is similar to V~I. This suggests that Ti~I is depleted in the observable atmosphere relative to the solar abundance assumed when calculating $\Psi$. We estimate the Ti~I abundance to be at least 10$\times$ below the solar abundance value for its detectability to fall below our threshold.

\section{Summary and Discussion} \label{sec:discussion}

We report new detections of V~I and Cr~I in the atmosphere of WASP-121~b at 4.5 $\sigma$ and 3.6 $\sigma$, respectively, for the first time in an exoplanet. We also confirm previous detections of Fe~I in the visible and Fe~II in UV. Based on their weighted absorption strengths, we infer the neutral species (Fe~I, V~I, and Cr~I) to pertain to the lower atmosphere and the ionic Fe~II to be in the upper atmosphere. Our study also demonstrates a viable metric to systematically predict the detectability of atomic species in UHJs.

These detections provide important constraints on the chemical and physical processes in the atmosphere of WASP-121~b. Transition metals (e.g., Ti, V, Fe), in both atomic and molecular forms, are known to be strong sources of optical opacity \citep{Fortney2008, Lothringer2020} which can influence radiative processes in UHJ atmospheres. The presence of Fe~I, V~I and Cr~I in the lower atmosphere may, therefore, provide the strong optical opacity required to explain the thermal inversion inferred in WASP-121~b \citep{Evans2017,Gibson2020, Lothringer2020}. Our detection of V~I argues for the thermal dissociation of VO in the observable atmosphere thereby explaining its non-detection in recent searches \citep{Merritt2020}. Our non-detection of atomic Ti despite its strong detectability suggests it is present as molecular TiO which may be condensed out at the terminator. 

These results also mark a significant development in our ability to find trace metals in the lower atmospheres of giant exoplanets. Both Cr and V are trace metals with solar abundances of 0.4 ppm and 0.008 ppm, respectively; two orders of magnitude lower than Fe (30 ppm). While previous searches suggested hints of these species in other UHJs \citep{Hoeijmakers2019,Cauley2019}, conclusive detections of the same remained elusive. Detecting such species is particularly important considering that they cannot be observed in planets cooler than $\sim$2000~K, including solar system giant planets. As such, they provide unique probes of the possibly rich diversity of refractory compositions of giant planets.

Our study highlights the power of high resolution spectroscopy to probe the metal compositions of UHJs, down to temperatures of $\sim$2000 K. The detections of five metal species (Na, Mg, Fe, Cr, V) in WASP-121b, which is at the cooler end of UHJs, implies that a wide range of UHJs could be exquisite laboratories for studying a plethora of such elements. 

\acknowledgements

This research has made use of the services of the ESO Science Archive Facility. The study was based on observations collected at the European Southern Observatory under ESO programme 0100.C-0750(C). We thank the anonymous referee for their helpful comments.

\bibliographystyle{aasjournal}




\end{document}